\def\beg{\begin{equation}}
\def\eeq{\end{equation}}
\documentstyle[12pt]{article}
\begin{document}
\begin{center}
{\Large{\bf Microwave resonance in 2D Wigner crystals}}
\vskip0.95cm
{\bf Keshav N. Shrivastava}
\vskip0.25cm
{\it School of Physics, University of Hyderabad,\\
Hyderabad  500046, India}
\end{center}
Recent experimental measurements of the microwave absorption in the frequency range 1- 4 GHz show peaks at several values of the filling factor, $\nu=nh/eB$. Since the filling factor multiplies the magnetic field, it is similar to the Lande's splitting factor, $g$. By using both signs in the $j=l\pm s$, we obtain a formula for the splitting factor which is equivalent to obtaining the filling factor. The experimentally observed values of 6/5, 4/3, 7/5, 8/5, 5/3, 7/3, and 5/2 are in agreement with our formula provided that we propose ``electron clusters". The clusters have ${\it l}$ and $s$ values and vary in size. It is possible to interpret the filling factor as a fractional charge. When charge becomes zero, the transverse resistivity diverges so that large resistivity belongs to an insulator which can be called a Wigner crystal.
\vskip1.0cm
Corresponding author: keshav@mailaps.org\\
Fax: +91-40-2301 0145.Phone: 2301 0811.
\vskip1.0cm

\noindent {\bf 1.~ Introduction} 

     Recently, a very important experiment has been performed by Chen et al[1]. It measures the microwave absorption in a device made of GaAs/AlGaAs. The size of the quantum well is 30 nm. The electron concentration is $n=3.0\times 10^{11} cm^{-2}$, the temperature is $\sim 50$ mK and the microwave frequency is $\sim 1-4$ GHz. The real part of the conductivity as a function of magnetic field shows dips whenever the filling factor becomes 4/3, 5/3, 7/3, 5/2, 6/5, 7/5 and 8/5.

     In this note, we wish to interpret the fractions at which there are dips in the absorption. We report that these fractions indicate the formation of ``electron clusters", of well defined angular momenta. At the point of zero charge, the resistivity diverges which creates a region of large resistivity. This region is called insulator or Wigner crystal.

\noindent{\bf 2.~~Theory}

     In recent years, there has been considerable interest in the problem of effective charge of a quasiparticle. Since charge appears  in the Bohr magneton, it can be derived in terms  of angular momenta operators. In this connection we have reported[2-11] that the effective charge of a quasiparticle is given by,
\beg
{1\over 2}g_{\pm}={e_{eff}\over e} = {{\it l}+{1\over 2}\pm s\over 2{\it l}+1}
\eeq

The details of the mathematical derivation are given in a book[8]. First of all, we propose that there are clusters of electrons which have well defined values of ${\it l}$ and $s$. In 
\vskip0.25cm
\begin{center}
{\bf Table 1:Calculated values of the filling\\
 factor for positive sign of spin\\
 for given values of ${\it l}$ and $s$. }\\
\begin{tabular}{cccc}
\hline

S.No. & ${\it l}$ & $S$   & $\nu_+$\\
\hline
1     &    0      &  2    &   5/2\\
2     &    1      &  5/2  &   4/3\\
3     &    1      &  7/2  &   5/3\\
4     &    1      &  11/2 &   7/3\\
5     &    2      &  7/2  &   6/5\\
6     &    2      &  9/2  &   7/5\\
7     &    2      &  11/2 &   8/5\\
8     &    2      &   4   &   13/2\\
9     &    2      &   5   &   15/2\\
\hline
\end{tabular}
\end{center}

\noindent that case, we can substitute some selected values of ${\it l}$ and $s$ in the above formula and obtain the effective charge of a quasiparticle. We substitute ${\it l}$
=0,1, 2, etc and $S$=5/2, 7/2, 11/2, ..., 1, 2, 3, ... etc. For a given value of $S$, the projections along the magnetic field have values $M_s$=-S, -(S-1), ...(S-1), S. For example, for $S$ =3/2, the $M_s$ values are $\pm$ 3/2, $\pm$ 1/2 and for S=1/2, the $M_s$ values are $\pm$ 1/2. The large $M_s$ transitions are suppressed and only -1/2 to +1/2 transition is visible. This type of suppression is known in liquids due to correlations or rotations and in glasses due to random distribution. If only -1/2 to 1/2 transition occurs, the transition energy will be ${1\over 2}g_{\pm}\mu_BH$. The Hamiltonian is simply $g\mu_BH.S$. The transitions show up in the resistivity because the later depends on the response function. Whenever the microwave frequency $h\nu$ becomes equal to the transition energy we expect a dip in the resistivity, i.e., at
\beg
h\nu_m={1\over 2}g_{\pm}\mu_BH
\eeq
There are new resonances, which occur at the fractional filling factor,
\beg
\nu_{\pm}={1\over 2}g_{\pm}
\eeq
Usually, the Landau levels are given by the cyclotron frequency,
\beg
E_L=(n+{1\over 2})\hbar \omega_c.
\eeq
Here
\beg
n=1
\eeq
\beg
\hbar\omega_c=h\nu_m=g\mu_BH
\eeq
where $\nu_m$ is the microwave frequency. The usual cyclotron resonance is given by (6) with no possibility of splitting. There is only one cyclotron frequency but now in the new theory there are two resonances,
\beg
h\nu_m=\hbar{1\over 2}g_+\omega_c
\eeq
and
\beg
h\nu_m=\hbar {1\over 2}g_-\omega_c
\eeq
The physical phenomenon is also slightly different because there are four eigen values,
\beg
E_1= {1\over 2}g_-\mu_BH(-{1\over 2})
\eeq
\beg
E_2= {1\over 2}g_-\mu_BH(+{1\over 2})
\eeq
\beg
E_3= {1\over 2}g_+\mu_BH(-{1\over 2})
\eeq
\beg
E_4={1\over 2}g_+\mu_BH(+{1\over 2}).
\eeq
The transition $E_4-E_3=h\nu_m$ is similar to ESR but new transitions are predicted. Some times, the magnetic field is high so that the levels containing $g_-$ are not populated so that $E_1$ and $E_2$ do not exist. Otherwise, more than ESR transitions are predicted.

     The resistivity is of the form $\rho_{xy}$=$h/ie^2$. When i=0, the charge becomes zero and the resistivity diverges. When the resistivity
becomes large, insulator is formed. This insulating state is some times called the ``Wigner crystal". There is a quasiparticle with spin 1/2 and zero charge. 
     
\noindent{\bf3.~~ Experiment}.

     The experimental measurements of microwave absorption have been performed by Chen et al [1]. The absorption is found at,
\beg
\nu = 6/5, 4/3, 7/5, 8/5, 5/3, 7/3, 5/2.
\eeq
These experimentally measured values are the same as the calculated values of Table 1. What is happening is that there are ``electron clusters". Every cluster has a definite value of the spin and there is an orbital angular momentum quantum number. Here, the Landau level
quantum number is $n=1$. At very low temperatures the excited states are not populated so that the experimental values are exactly as calculated. At higher temperatures, the excited states get mixed in the ground state due to the spin-orbit interaction so that the calculated values may differ from the measured values unless the spin-orbit interaction is included in the calculation. At higher temperatures, there is the electron-phonon interaction so that the experimental values deviate from the tabulated values.

\noindent{\bf 4.~~Conclusions}

     In GaAs/AlGaAs heterostructures microwaves are absorbed according to new resonances which occur due to both the positive as well as negative values of the spin. This phenomenon is slightly different from electron spin resonance which involves absorption by single flip of spin. There are electron clusters of definite values of ${\it l}$ and $s$. We predict the filling factors by using ${\it l}$ and $s$. Since the temperature of measurement is very low, the populations in the excited states are minimized and the electron-phonon interaction is
largely reduced so that the theoretically predicted values are the same as those measured.

\vskip1.25cm
\noindent{\bf5.~~References}
\begin{enumerate}
\item Y. Chen, R. M. Lewis, L. W. Engel, D. C. Tsui, P. D. Ye, L. N. Pfeiffer and K. W. West, Phys. Rev. Lett. {\bf 91},016801(2003).
\item K. N. Shrivastava, Phys. Lett. A {\bf 113}, 435 (1986).
\item K. N. Shrivastava, Mod. Phys. Lett. B {\bf 13}, 1087 (1999).
\item K. N. Shrivastava, Mod. Phys. Lett. B {\bf 14}, 1009 (2000).
\item K. N. Shrivastava, cond-mat/0212552, cond-mat/0303309, cond-mat/0303621
\item K. N. Shrivastava in Frontiers of fundamental physics 4, edited by B. G. Sidharth and M. V. Altaisky, Kluwer Academic/Plenum Pub. New York 2001.
\item K. N. Shrivastava, Superconductivity: Elementary Topics, World Scienctific, New Jersey, London, Singapore, 2000.
\item K.N. Shrivastava, Introduction to quantum Hall effect,\\ 
      Nova Science Pub. Inc., N. Y. (2002).
\item K. N. Shrivastava, cond-mat/0302610.
\item K. N. Shrivastava, Natl. Acad. Sci. Lett. (Ibdia), {\bf 9}, 145 (1986).
\item K. N. Shrivastava, Natl. Acad. Sci. Lett. (India), (2003).
\end{enumerate}
\vskip0.1cm

Note: Ref.8 is available from:\\
 Nova Science Publishers, Inc.,\\
400 Oser Avenue, Suite 1600,\\
 Hauppauge, N. Y.. 11788-3619,\\
Tel.(631)-231-7269, Fax: (631)-231-8175,\\
 ISBN 1-59033-419-1 US$\$69$.\\
E-mail: novascience@Earthlink.net

\end{document}